\pdfoutput=1
\documentclass[10pt,conference]{IEEEtran}
\usepackage{graphicx}
\IEEEoverridecommandlockouts

\usepackage{booktabs}
\usepackage{enumitem}
\usepackage{xcolor}
\usepackage{url}
\usepackage{listings}
\usepackage[most]{tcolorbox}
\usepackage{xcolor}
\usepackage[table]{xcolor}
\usepackage{colortbl}
\usepackage{array}
\usepackage{listings}
\usepackage{microtype}
\usepackage{float}
\usepackage{mdframed}
\usepackage{parskip}
\usepackage{xspace}
\usepackage{hyperref}
\usepackage{balance}

\definecolor{lightbluebg}{RGB}{235, 245, 255}
\definecolor{borderblue}{RGB}{180, 205, 235}
\definecolor{greenheader}{HTML}{3D6B3D}
\definecolor{greenlight}{HTML}{EAF3EA}
\definecolor{greenwhite}{HTML}{F7FBF7}

\definecolor{redheader}{HTML}{7A2E2E}
\definecolor{redlight}{HTML}{FAEAEA}
\definecolor{redwhite}{HTML}{FDF6F6}

\definecolor{blueheader}{HTML}{2E4E6E}
\definecolor{bluelight}{HTML}{E8EFF7}
\definecolor{bluewhite}{HTML}{F4F7FB}

\definecolor{purpleheader}{HTML}{4A2E6E}
\definecolor{purplelight}{HTML}{EEE8F7}
\definecolor{purplewhite}{HTML}{F8F5FC}

\definecolor{traceheader}{HTML}{2E5555}
\definecolor{tracelight}{HTML}{E8F4F4}
\definecolor{tracewhite}{HTML}{F4FAFA}

\newtcolorbox{systempromptbox}{
  colback=lightbluebg,
  colframe=borderblue,
  arc=6pt,
  boxrule=0.8pt,
  left=10pt,
  right=10pt,
  top=8pt,
  bottom=8pt,
  enhanced,
  before skip=4pt,   
  after skip=4pt,    
}

\newtcolorbox{rqanswer}[1]{
  colback=gray!10!white,
  colframe=gray!50!black,
  coltitle=white,
  fonttitle=\bfseries\small,
  title={#1},
  sharp corners,
  boxrule=0.8pt,
  titlerule=0pt,
  toptitle=3pt,
  bottomtitle=3pt,
  left=6pt, right=6pt, top=4pt, bottom=4pt
}
\hypersetup{
  colorlinks=true,
  linkcolor=blue,
  citecolor=blue,
  urlcolor=blue
}

\newcommand{\ignore}[1]{}

\newcommand{\toolname}{TRACE\xspace}

\newcounter{appsec}

\begin{document}
\title{Measuring LLM Trust Allocation Across Conflicting Software Artifacts}

\author{
\IEEEauthorblockN{Noshin Ulfat}
\IEEEauthorblockA{noshin.ulfat@utdallas.edu\\
\textit{University of Texas at Dallas}\\
Texas, USA}
\and
\IEEEauthorblockN{Ahsanul Ameen Sabit}
\IEEEauthorblockA{ahsanulameen.sabit@utdallas.edu\\
\textit{University of Texas at Dallas}\\
Richardson, Texas, USA}
\and
\IEEEauthorblockN{Soneya Binta Hossain}
\IEEEauthorblockA{sbhossain@utdallas.edu\\
\textit{University of Texas at Dallas}\\
Texas, USA}
}
\maketitle

\begin{abstract}

LLM-based software engineering assistants often reason over multiple artifacts, including code, documentation, signatures, and tests, even when those artifacts are incomplete or \textit{mutually inconsistent}. Existing evaluations primarily measure final outputs, leaving unclear whether a model recognized unreliable evidence, identified the faulty source, or prioritized the appropriate artifact.

We introduce \toolname{}, a controlled method for evaluating how LLMs assess and prioritize conflicting software artifacts. \toolname{} constructs paired clean and perturbed versions of real-world Java method bundles by injecting known faults into the documentation, implementation, or both while holding the remaining artifacts fixed. Models then assess artifact quality, detect and localize inconsistencies, and rank the available sources by reliability. Using 22,339 valid responses from seven LLMs on 456 method bundles, we find that quality penalties are generally localized to the perturbed artifact and increase with fault severity. However, models exhibit a consistent source-origin asymmetry: they detect documentation faults at 67--94\% and explicit documentation--implementation contradictions at 50--91\%, but detection falls by 21--43 percentage points when only the implementation changes while documentation remains intact. Models also struggle to deprioritize faulty implementations, and confidence provides little separation between correct and incorrect judgments for six of seven models. These results show that current LLMs are not symmetric integrators of software evidence: they audit natural-language specifications more reliably than subtle implementation behavior. \textit{\toolname{} provides a controlled method for exposing this failure mode before LLMs are used in correctness-critical software engineering workflows.}

\end{abstract}

\section{Introduction}
\begin{figure*}[t]
  \centering
  \includegraphics[width=\linewidth]{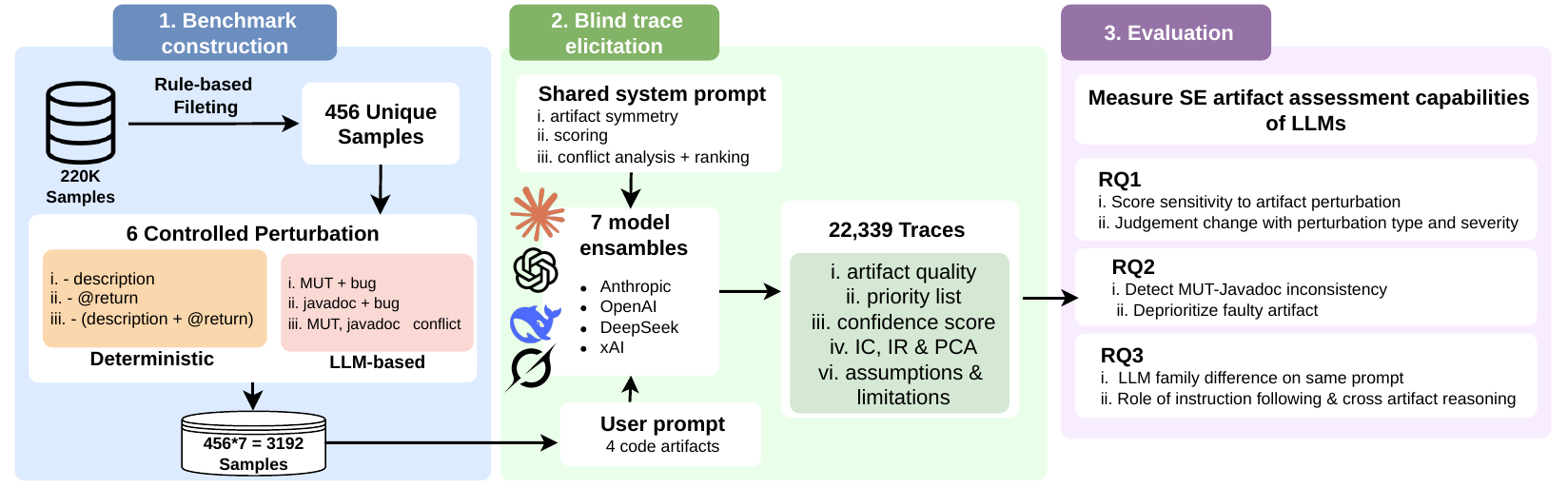}
  \caption{Overview of \toolname Pipeline.}
  \label{fig:pipeline}
\end{figure*}

A software engineering assistant can produce a correct output while relying on the wrong evidence. Consider a method whose Javadoc specifies one behavior while its implementation encodes another. An LLM may still generate a correct test oracle, patch, or review comment---perhaps even one that passes the available tests. The output alone, however, does not reveal whether the model detected the conflict, identified the unreliable artifact, or simply followed the source it tends to privilege. This distinction is critical: a model that succeeds by trusting the convenient artifact may fail silently when the source of the inconsistency changes.

This problem arises throughout LLM-based software engineering. Test generation, patching, requirements analysis, summarization, and code review all require models to reason over bundles of code, documentation, signatures, tests, and contextual instructions that may be incomplete, stale, or mutually inconsistent~\cite{Hou2024LLM4SE,Fan2023LLM4SE}. Yet most evaluations focus on final-output correctness. \textit{They therefore conflate two capabilities: determining which evidence is reliable and producing an answer from that evidence}. As a result, a high task score cannot distinguish a model that reconciles conflicting artifacts from one that reaches the same answer by defaulting to an unreliable source.

Prior work shows that artifact quality affects downstream performance, but does not directly evaluate this evidence-selection problem. Studies of test-oracle generation show that documentation materially influences generated assertions and that learned or LLM-based generators remain vulnerable to noisy context, false positives, and generalization gaps~\cite{Doc2OracLL,HossainNeural2023,togll}. Complementary work detects inconsistencies between documentation and program behavior or verifies consistency across multiple artifacts~\cite{Lee2025METAMON,Xu2025DocPrism,dietrich2025llm}. These approaches answer important but different questions: downstream-generation studies ask whether an output is correct, while inconsistency detectors ask whether artifacts disagree. They do not jointly measure whether a model assesses each artifact's quality, localizes the unreliable source, prioritizes competing evidence, and calibrates its confidence. Moreover, they generally do not test \emph{source-origin symmetry}: whether a model responds consistently when an equivalent conflict originates in documentation rather than implementation. \textit{Thus, we lack a controlled way to determine whether LLMs genuinely integrate heterogeneous software evidence or systematically privilege one artifact class}.

Testing provides a precise setting for closing this gap. A unit-test task combines artifacts with distinct evidentiary roles: the method signature constrains types and exceptions, the Javadoc expresses documented intent, the method under test (MUT) encodes current behavior, and the test prefix defines the execution scenario. None is inherently authoritative: documentation may be stale, code may be faulty, and test context may be incomplete or partial. At the same time, these artifacts can be perturbed independently while the remainder of the bundle is held fixed. Testing therefore provides both a correctness-relevant application and a controlled environment for isolating how models respond to missing, degraded, and contradictory evidence.

We use \emph{artifact-level trust} to denote a model's allocation of reliability across the artifacts supplied for a task. We measure it through four observable judgments: \textit{per-artifact quality assessment}, \textit{inconsistency detection}, \textit{affected-artifact attribution}, and \textit{source prioritization}, together with reported confidence. These outputs reveal not only whether a model notices a problem, but also where it locates the problem and which evidence it would rely on downstream.

We introduce \toolname{} in Figure~\ref{fig:pipeline}, a model-neutral software engineering framework for evaluating artifact-level trust under controlled perturbations. In our Java testing instantiation, \toolname{} creates paired clean and perturbed method bundles by modifying the Javadoc, the MUT, or both while holding other artifacts fixed. Under a blind protocol, models assess artifact quality, detect and localize conflicts, rank sources by reliability, and report confidence; explicit provenance isolates the effect of each perturbation. Our framework is applicable to requirements, code, tests, documentation, patches, and agent context, with perturbations tailored to workflow-specific risks. \toolname{} therefore supports comparative research and predeployment auditing by exposing over-trusted or under-scrutinized artifacts, guiding model and prompt selection, identifying where human or programmatic checks are needed, and enabling repeatable regression testing as models evolve.

We evaluate seven LLMs using 22,339 valid responses over 456 method bundles from 25 real-world Java systems. Quality penalties are generally \textit{localized} to the perturbed artifact and increase with fault severity, showing that models capture more than binary fault presence. This sensitivity, however, is strongly \textit{asymmetric}. \textit{Models detect documentation faults at 67--94\% and explicit Javadoc--MUT contradictions at 50--91\%, but detection falls by 21--43 percentage points when only the implementation changes while documentation remains unchanged}. Models also \textit{struggle} to \textit{deprioritize faulty implementations}, and confidence meaningfully separates correct from incorrect judgments for \textit{only} one of seven models. These findings show that current LLMs are \textit{not symmetric} integrators of software evidence: \textit{they audit natural-language specifications more reliably than subtle implementation behavior}.

This asymmetry matters beyond testing. Code review, requirements analysis, patch generation, and maintenance all require deciding which artifact remains credible when evidence conflicts. \textit{By making that decision observable and measurable, \toolname{} helps researchers characterize artifact-specific model behavior and helps practitioners expose hidden blind spots before deploying LLMs in correctness-critical workflows}.

In summary, this work makes contributions:
\begin{itemize}
    \item We formalize artifact-level trust as a software engineering evaluation target covering artifact quality, conflict detection and attribution, source prioritization, and confidence.

    \item We introduce \toolname{}, a model-neutral controlled-perturbation framework, instantiated as a provenance-annotated benchmark of 456 Java method bundles with clean, degraded, and contradictory variants.

    \item We evaluate seven LLMs and reveal severity-sensitive yet artifact-asymmetric judgments, including an implementation-only blind spot, weak deprioritization of faulty code, and poor confidence calibration.

    \item We release reproducible artifacts and distill practical guidance to support replication, reuse, and future benchmark extensions.
\end{itemize}

\section{Approach}

Figure~\ref{fig:pipeline} summarizes the \toolname{} pipeline. The approach has three steps:
(1) define the artifact-level trust task, (2) construct aligned base and
perturbed artifact bundles, and (3) elicit structured trust judgments
under a blind protocol. Since each perturbation modifies only selected
artifacts while holding the method and remaining context fixed, changes
in model judgments can be attributed to the injected perturbation.

\subsection{Task Definition}
\label{sec:task_definition}

\paragraph{Input bundle.}
For method $i$ and variant $v$, the model receives
\[
s_i^v =
\left\langle M_i^v,\sigma_i,J_i^v,\tau_i \right\rangle,
\]
where $M_i^v$ is the method under test (MUT), $\sigma_i$ is its
signature, $J_i^v$ is the Javadoc, and $\tau_i$ is the test prefix.
Only the MUT and Javadoc vary across perturbations; the signature and
test prefix remain fixed. A reference assertion $O_i$ is available for
419 of the 456 bundles and is used only for \textit{perturbation construction
and sanity checking}. It is neither shown to the model nor treated as a
prediction target.

\paragraph{Structured trust trace.}
\label{sec:formula}
We define \emph{artifact-level trust} as a model's explicit assessment,
comparison, and prioritization of the supplied artifacts. For model $m$,
the response is a structured trace
\[
r_i^{v,m} =
\left\langle A,C,P,H,\psi,c \right\rangle,
\]
where:
\begin{itemize}[leftmargin=1.2em]
    \item $A$ contains quality assessments for the Javadoc, signature,
    MUT, test prefix, and overall bundle. Each assessment includes a
    score in $[0,1]$, a categorical label, and supporting evidence.

    \item $C$ contains three conflict views: pairwise conflict analysis
    ($C_{\mathrm{PCA}}$), an explicitly identified conflict list
    ($C_{\mathrm{IC}}$), and a consolidated inconsistency report
    ($C_{\mathrm{IR}}$). The schema also records detected anomalies.

    \item $P$ is a total reliability ordering over
    $\{\textsc{Javadoc},\textsc{Signature},\textsc{MUT},
    \textsc{TestPrefix}\}$, with confidence for each rank.

    \item $H$ records the most defensible behavioral interpretation of
    the bundle, $\psi$ records assumptions and limitations, and
    $c\in[0,1]$ is the overall confidence.
\end{itemize}

Here, \emph{trace} denotes a structured record of \emph{observable} and \emph{measurable} judgments,
not a model's hidden chain of thought. Later analyses derive detection
indicators from $C_{\mathrm{PCA}}$, $C_{\mathrm{IC}}$, and
$C_{\mathrm{IR}}$, together with their Union and Majority aggregates.

\subsection{Controlled Benchmark Construction}
\label{sec:benchmark_construction}

\subsubsection{Base-Sample Curation}

We derive the benchmark from OE25 dataset, containing 223,557 input samples
from 56 modules across 25 real-world Java systems~\cite{HossainNeural2023, togll}.
To reduce pre-existing noise, we retain only methods that
(1) contain 8--60 non-comment executable lines,
(2) exhibit substantive behavior, defined as at least two control-flow constructs
or at least four assignments and four method calls, and
(3) are not constructors, entry points, trivial accessors or wrappers,
builder chains, extraction artifacts, or methods with inline comments
that could provide unintended cues.

We further require
(1) English Javadocs with substantive prose longer than 15 characters,
(2) an \texttt{@return} tag for non-void methods and at least one
\texttt{@param} tag for parameterized methods, and
(3) a test prefix of at least 20 characters. We exclude inherited or
tag-only documentation and duplicate MUT--Javadoc pairs, yielding 456
curated method bundles.\\

\subsubsection{Perturbation Variants}
\label{subsec:perturbation_variants}

For each base bundle, we construct six aligned variants, summarized in
Table~\ref{tab:variants}.

\begin{table}[H]
\centering
\footnotesize
\setlength{\tabcolsep}{4pt}
\renewcommand{\arraystretch}{1.15}
\caption{\toolname{} perturbation variants.}
\label{tab:variants}
\resizebox{\columnwidth}{!}{%
\begin{tabular}{@{}lll@{}}
\toprule
\textbf{Variant} & \textbf{Operation} & \textbf{Severity} \\
\midrule
\textsc{NoDesc}
    & Remove Javadoc prose; retain tags
    & --- \\

\textsc{NoReturn}
    & Remove \texttt{@return}, when present
    & --- \\

\textsc{NoDescReturn}
    & Remove prose and \texttt{@return}
    & --- \\

\textsc{DocBug}
    & Perturb Javadoc; retain the base MUT
    & H/N/S \\

\textsc{MutBug}
    & Perturb MUT; retain the base Javadoc
    & H/N/S \\

\textsc{Conflict}
    & Create a Javadoc--MUT conflict by changing
      Javadoc, MUT, or both
    & H/N/S \\
\bottomrule
\end{tabular}}
\end{table}

The removal variants are deterministic and exactly reproducible. The
generated variants use structured mutation prompts with three severity
tiers: \emph{heavy} faults create explicit contradictions,
\emph{normal} faults occur under specific inputs or boundary conditions,
and \emph{subtle} faults introduce small corner-case deviations requiring
close behavioral comparison. Each generated family is balanced across
the three tiers, and \textsc{Conflict} additionally balances three fault
origins: Javadoc only, MUT only, and both artifacts.

All generated mutations preserve the method signature and test prefix,
remain natural in form, and avoid comments or markers that disclose the
perturbation. MUT variants must compile. We manually inspect each
generated variant for correct realization and severity, and retain
provenance metadata for the modified artifact, fault category, severity,
change, and expected behavioral conflict.

The base and six variants yield $456 \times 7 = 3{,}192$ aligned inputs.
Variants from the same method are treated as paired observations. The
deterministic Javadoc removals apply to the description, return tag, or
both; description removal applies to all 456 bundles, and return-tag
removal applies to 384. The generated variants produce 456 samples each,
balanced across heavy, normal, and subtle tiers (152 per tier).

Fault distributions follow real-world bug frequencies rather than
uniform sampling. MUT faults are mainly logic faults (69\%), followed by
boundary faults (16\%), API misuse (8\%), and null-check faults (7\%).
Javadoc faults mirror this distribution with wrong behavior (68\%),
wrong return information (17\%), wrong parameter information (11\%),
and missing information (4\%). Javadoc--MUT contradictions are evenly
split across Javadoc-only, MUT-only, and both-artifact injections
(152 each), spanning return value (26\%), logic (26\%), null handling
(22\%), exception behavior (17\%), and side effects (10\%). Our
\toolname{} artifact scripts provide the full details.

To validate perturbation quality, two annotators with Java and
software-testing experience independently labeled
$456 \times 3 = 1{,}368$ generated samples for modified artifact,
unreliable artifact, contradiction validity, and severity. Disagreements
were resolved through annotators judgment.

\begin{figure*}[!t]
  \centering
  \includegraphics[width=\linewidth]{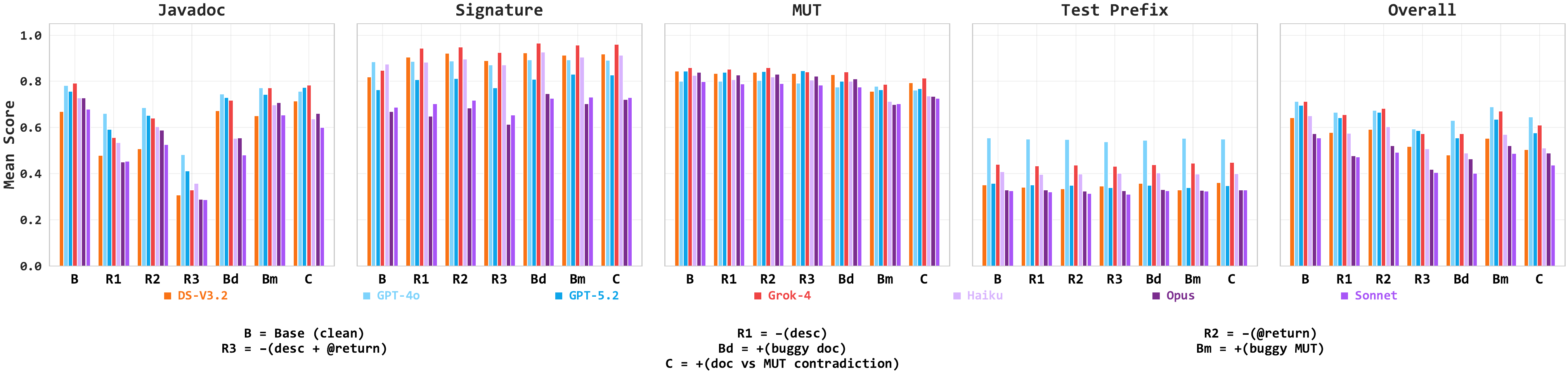}
 \caption{Mean artifact-quality scores by model and dataset variant for
Javadoc, signature, MUT, test prefix, and overall bundle. The figure shows
model-specific score calibration and the artifacts receiving the largest
quality penalties under each perturbation.}
  \label{fig:rq1_mean_scores_by_dataset}
\end{figure*}

\subsection{Trace Generation}
\label{sec:trace_elicitation}

\subsubsection{Prompt Protocol}

All models receive the same system prompt and user-template structure.
The protocol has three invariants: (1) \emph{artifact-symmetric}: no
artifact is marked authoritative, and models are instructed not to
default to either Javadoc or implementation; (2) \emph{blind}: the prompt
contains only the four artifacts and omits the dataset variant, modified
artifact, and perturbation severity; and (3) \emph{schema-constrained}:
each response must be a single JSON object following the trace schema in
Section~\ref{sec:task_definition}. These controls make judgments
comparable across models and variants.\\

\subsubsection{Execution and Schema Validation}

For each $(m,i,v)$ tuple, the pipeline loads the artifact bundle,
instantiates the fixed prompt, queries model $m$, parses and
schema-validates the JSON response, and stores the result incrementally.
The implementation is parallel and resumable, with endpoint-specific
retry and backoff policies that do not alter the prompt.

Validation checks structural compliance only; semantic correctness is
evaluated later against perturbation provenance. Each record stores the
input bundle, optional reference assertion, perturbation metadata,
structured trace, model identifier, and confidence scores. The canonical
run yielded 22,339 valid traces from 22,344 model--input trials
(99.98\%).

\section{Experimental Study}
\label{sec:experimental_study}

We study three questions. \textbf{RQ1} asks whether models localize input degradation and scale quality judgments with severity. \textbf{RQ2} asks whether they detect Javadoc--MUT conflicts, identify and deprioritize the faulty artifact, and calibrate confidence. \textbf{RQ3} asks whether robustness to subtle conflicts reflects semantic code understanding. Together, these questions evaluate sensitivity, actionability, and robustness.\\

\paragraph{Model Selection.}
We evaluate seven models:
Claude Opus 4.6, Sonnet 4.6, and Haiku 4.5; GPT-5.2 Chat and GPT-4o;
DeepSeek-V3.2-Speciale; and Grok 4 Fast Reasoning~\cite{anthropic2026claudeopus46,anthropic2026claudesonnet46,anthropic2025claudehaiku45,openai2025gpt52,openai2024gpt4o,deepseek2025v32speciale,xai2025grok4fast}.
The suite spans providers and capability tiers: the Claude models support
a within-family comparison, GPT-4o provides a prior-generation reference,
and DeepSeek and Grok broaden provider coverage.
This selection was informed by LiveBench coding and reasoning
results~\cite{white2025livebench}. All models use the same harness,
prompts, and temperature of 0.0. The released weights for
DeepSeek-V3.2-Speciale also provide a non-proprietary path for
reproducing the protocol.

\subsection{RQ1: Input-Quality Sensitivity}
\label{sec:rq1}

An LLM assistant should recognize when its input evidence is unreliable.
RQ1 tests whether models identify \textit{which} artifact was degraded and \textit{whether}
their quality judgments scale with perturbation \textit{severity}.

\subsubsection{Experimental Setup}

RQ1 analyzes the assessment component $A$ of the structured trace
$r_i^{v,m}=\left\langle A,C,P,H,\psi,c \right\rangle$ defined in
Section~\ref{sec:formula}. From each trace, we extract
\[
A=\langle a_{\text{javadoc}},a_{\text{signature}},a_{\text{mut}},
a_{\text{testprefix}},a_{\text{overall}}\rangle .
\]
We first report mean scores by model and variant. We then measure
sensitivity as the score change from the base:
\[
\Delta_{m,v,x}=\bar{a}_{m,v,x}-\bar{a}_{m,0,x},
\]
where $m$ is the model, $v$ is the perturbation, and $x$ is the assessed
artifact. Negative $\Delta$ values indicate lower perceived quality
relative to the base. Finally, for \textsc{DocBug}, \textsc{MutBug}, and
\textsc{Conflict} (discussed in Section~\ref{subsec:perturbation_variants}), we compare scores across subtle, normal, and heavy
perturbations. These analyses test whether $A$ captures both the \textit{location}
and \textit{severity} of input degradation.\\

\begin{figure*}[!t]
  \centering
  \includegraphics[width=\linewidth]{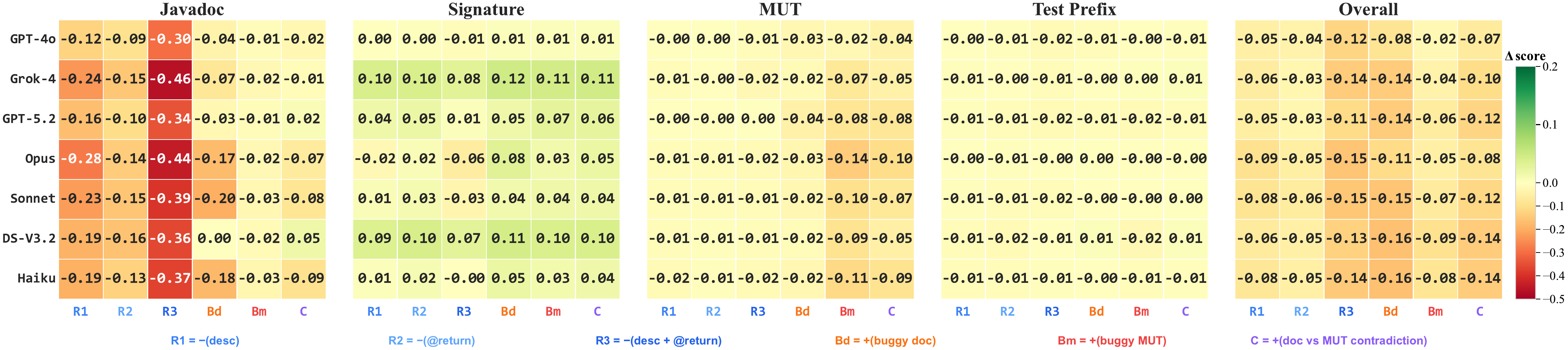}
  \caption{Quality-score change relative to the base variant,
$\Delta=\bar{a}_{m,v,x}-\bar{a}_{m,0,x}$, by model, perturbation, and
artifact dimension. Negative values indicate lower perceived quality;
concentrated negative values indicate localized sensitivity to the
perturbed artifact.}
  \label{fig:rq1_delta_from_base}
\end{figure*}

\begin{figure*}[!t]
  \centering
  \includegraphics[width=\linewidth]{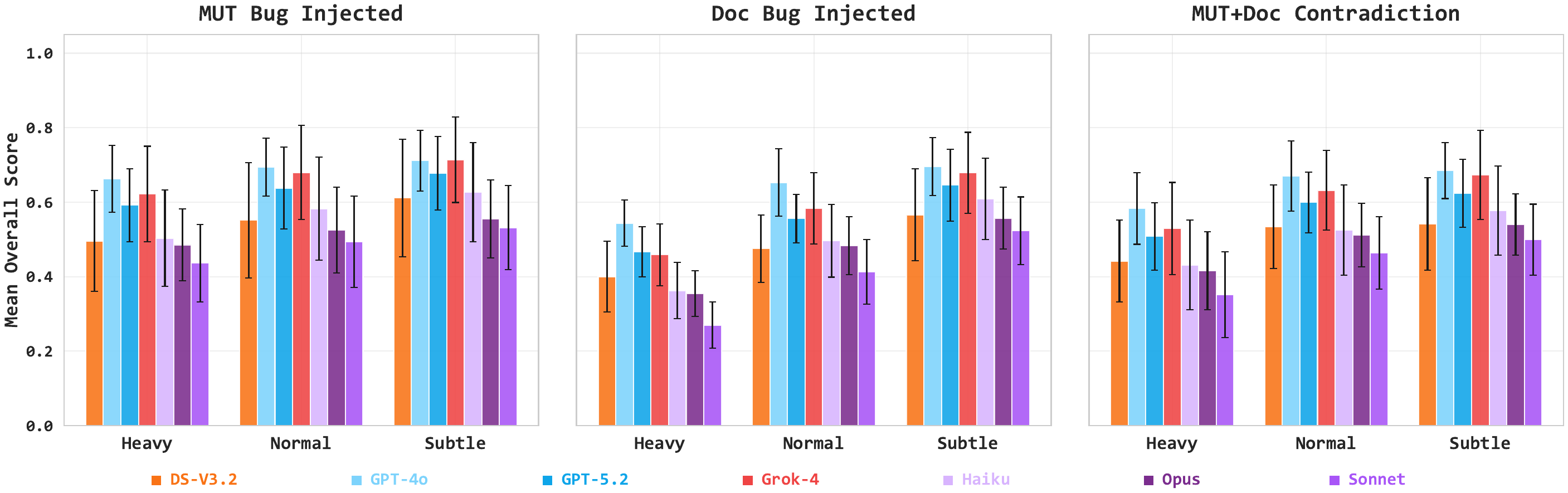}
\caption{Mean overall quality score by model and severity tier for
\textsc{DocBug}, \textsc{MutBug}, and \textsc{Conflict}; error bars show
$\pm$1 standard deviation. Wider heavy-to-subtle gaps indicate stronger
severity sensitivity.}
  \label{fig:rq1_severity_breakdown}
\end{figure*}

\subsubsection{Results}

\emph{RQ1 is positive, with an important qualification: models localize
and grade degradation, but sensitivity is artifact-dependent.}
Models use different score scales on the unperturbed base
(mean overall scores: 0.555--0.713), so we compare within-model deltas
rather than absolute scores
(Figure~\ref{fig:rq1_mean_scores_by_dataset}).

\emph{Penalties concentrate on the modified artifact.}
Removing both the Javadoc description and \texttt{@return} lowers
Javadoc scores by $0.300$--$0.463$, while MUT scores change by less than
$0.020$ (Figure~\ref{fig:rq1_delta_from_base}). This pattern indicates
artifact-specific sensitivity rather than uniform pessimism. However,
overall scores fall by only $0.109$--$0.155$, showing that models only
partially propagate local degradation to bundle-level risk.

\emph{The main pattern is a Javadoc--MUT asymmetry.}
Across all seven models, Javadoc faults receive larger penalties and
clearer severity separation than MUT faults. The heavy-to-subtle gap is
$0.152$--$0.253$ for Javadoc bugs, but only $0.049$--$0.123$ for MUT
bugs (Figure~\ref{fig:rq1_severity_breakdown}). Thus, quality scores
capture more than binary fault presence, but they are much less
discriminative for implementation degradation.

\emph{Implication.}
Artifact-quality scores can help localize degraded context, especially
degraded Javadocs. They under-signal implementation faults, however, and
should not be used as a stand-alone quality gate.

\begin{rqanswer}{RQ1 Findings}
LLMs localize and grade input degradation, but not symmetrically across
artifacts. Removing both the Javadoc description and \texttt{@return}
lowers Javadoc scores by $0.300$--$0.463$, while MUT scores change by
less than $0.020$. Scores also preserve severity ordering, but the
heavy-to-subtle separation is larger for Javadoc bugs
($0.152$--$0.253$) than for MUT bugs ($0.049$--$0.123$). Thus, models
are markedly \textbf{more sensitive to Javadoc degradation than to implementation
faults}.
\end{rqanswer}

\subsection{RQ2: Javadoc–MUT consistency detection}
\label{sec:rq2}

\begin{figure*}[!t]
  \centering
  \includegraphics[width=\linewidth]{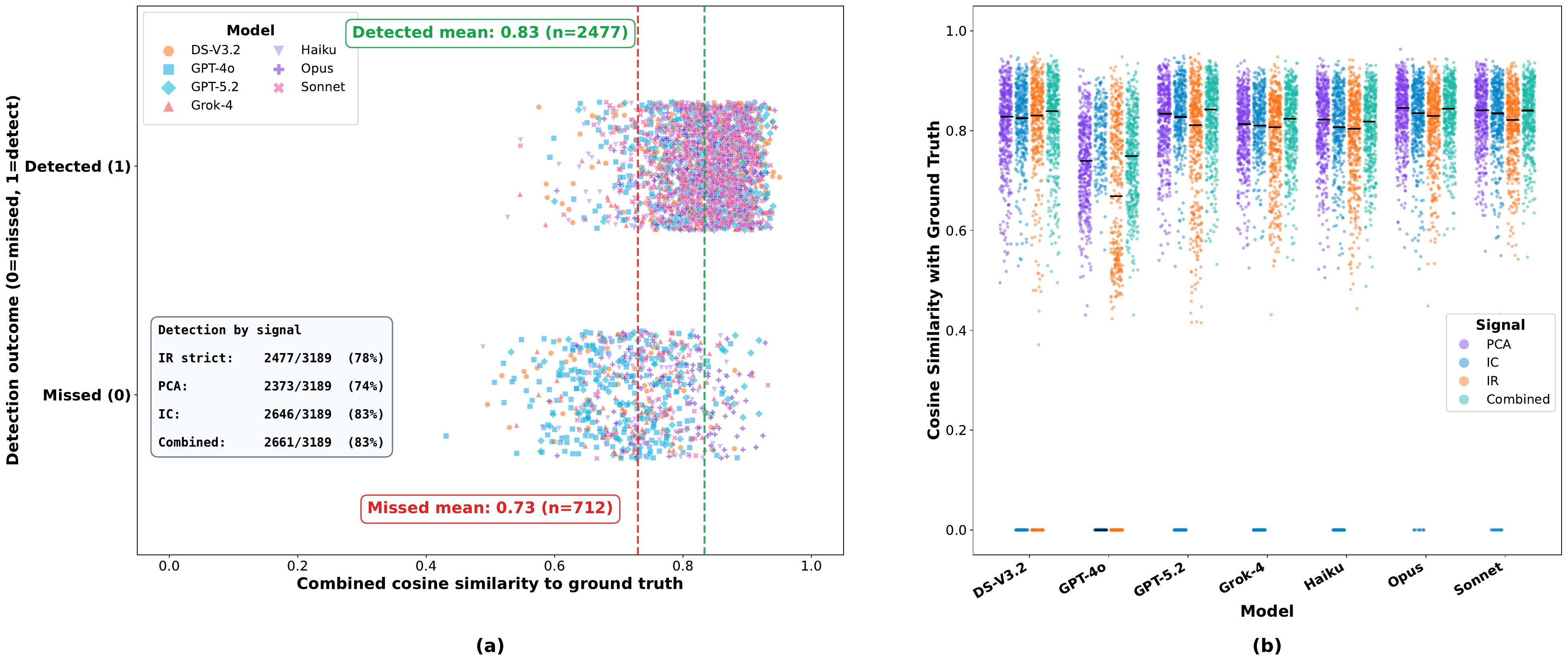}
  \caption{Description fidelity for \textsc{Conflict} traces, measured
as cosine similarity to ground-truth fault descriptions
($N=3{,}189$; embedding: \texttt{BAAI/bge-base-en-v1.5}). Panel~(a)
compares IR-strict detected and missed traces; detected traces have
higher similarity on average ($0.83$, $n=2{,}477$) than missed traces
($0.73$, $n=712$). Panel~(b) shows per-sample cosine distributions for
PCA, IC, IR, and their combined representation.}
  \label{fig:rq2_cosine_similarity}
\end{figure*}

\begin{figure*}[!t]
  \centering
  \includegraphics[width=0.5\linewidth]{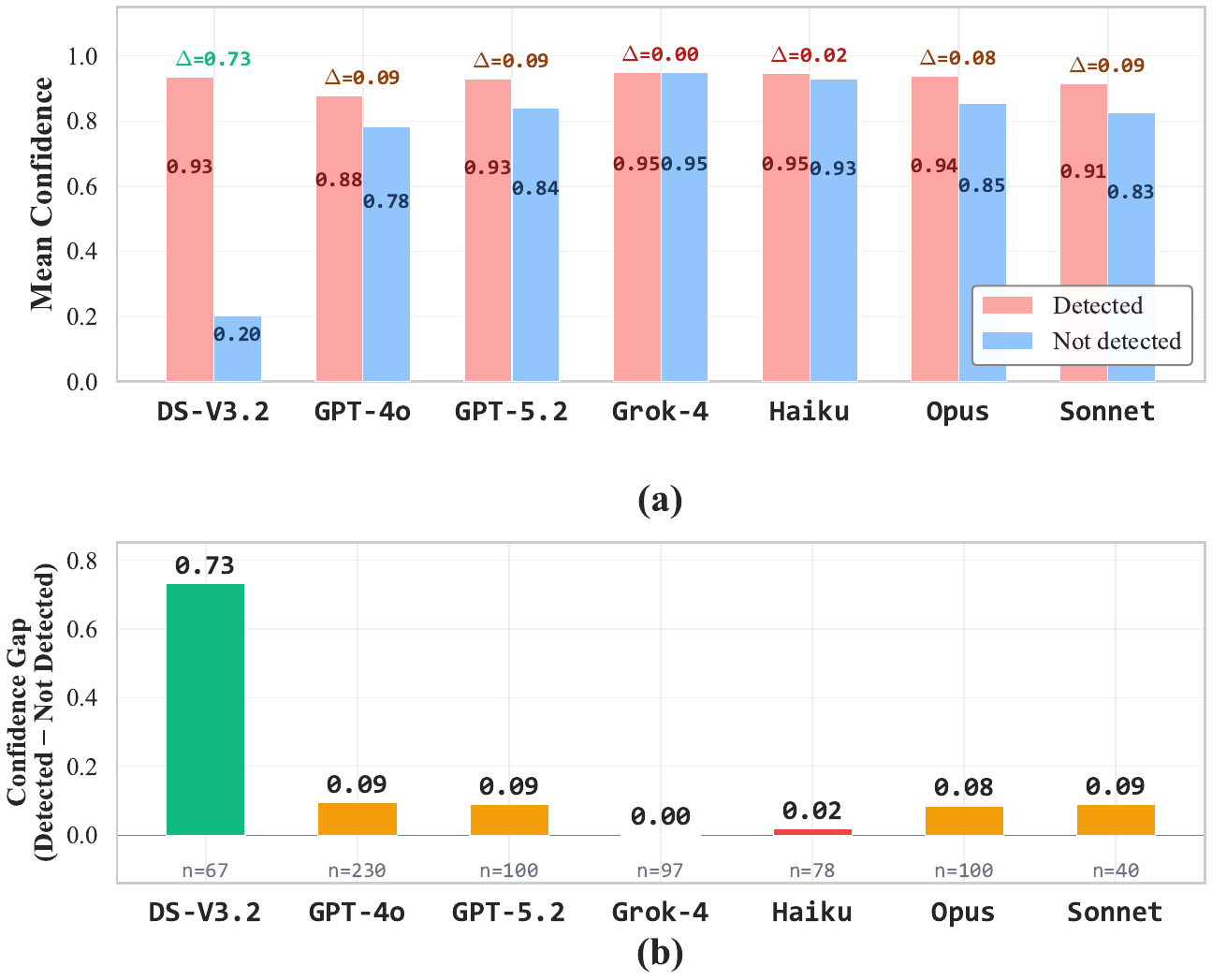}
  \caption{Confidence calibration on \textsc{Conflict} under IR-strict
evaluation. Panel~(a) compares mean confidence for detected and missed
traces. Panel~(b) shows the detected--missed confidence gap, with the
number of missed samples shown below each bar. DeepSeek-V3.2-Speciale
shows the largest separation ($0.73$); the other six models show little
separation.}
  \label{fig:rq2_confidence_calibration}
\end{figure*}

\begin{figure*}[!t]
  \centering
  \includegraphics[width=0.6\linewidth]{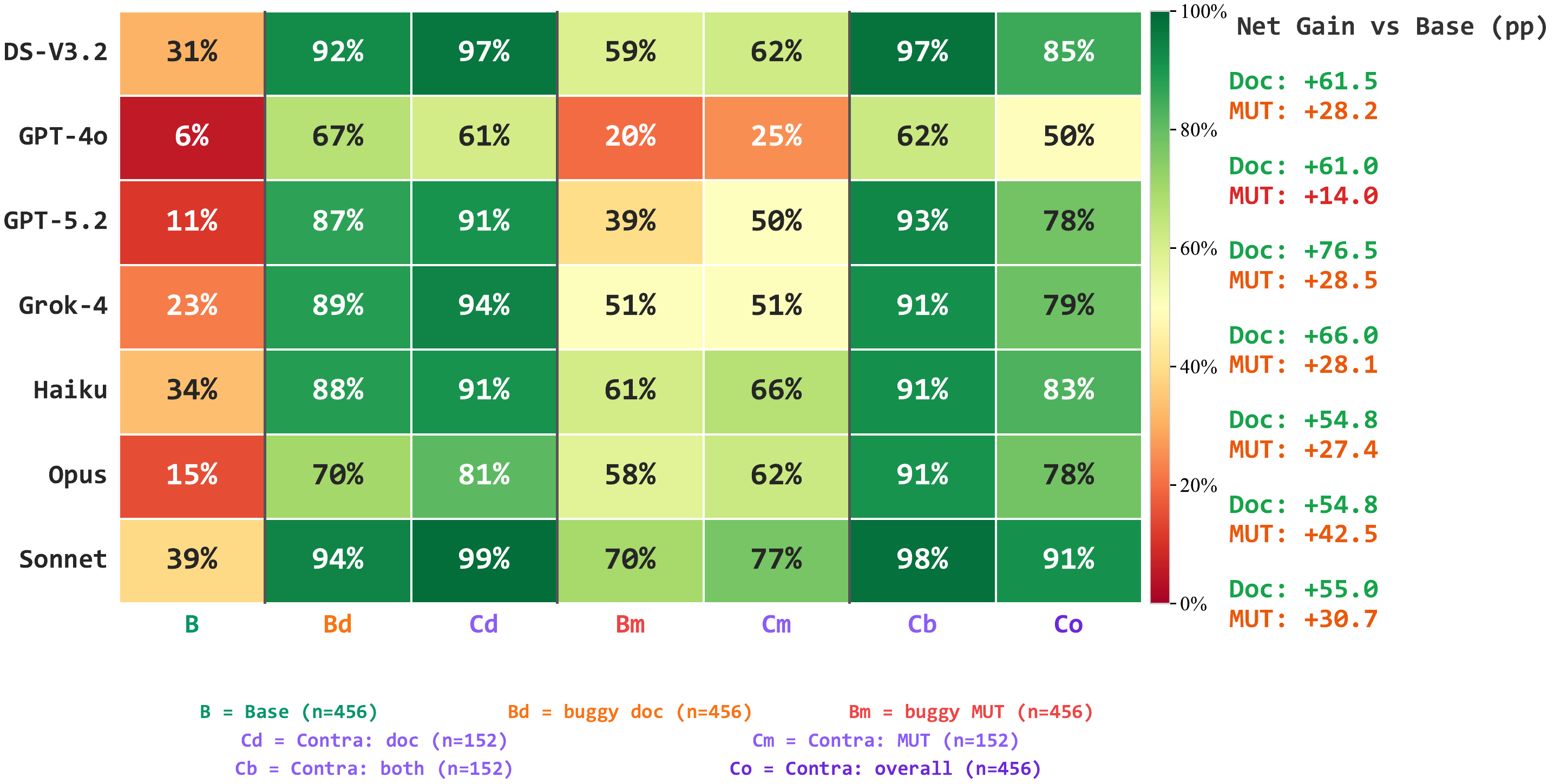}
  \caption{Javadoc-versus-MUT detection asymmetry under IR-strict
evaluation. Bars report net detection gain above the clean false-positive
baseline. Javadoc faults are detected more reliably than MUT faults
across all seven models: $+54.8$--$+76.5$~pp for \textsc{DocBug}
versus $+14.0$--$+42.5$~pp for \textsc{MutBug}.}
  \label{fig:rq2_doc_vs_code_asymmetry}
\end{figure*}

Lowering a quality score is not enough for an SE assistant to act safely.
When Javadoc and code disagree, the model must surface the conflict,
describe the mismatch, and adjust which artifact it trusts. RQ2 asks
whether models make these actionable judgments, and whether their behavior
depends on whether the fault originates in the Javadoc or the implementation.\\

\subsubsection{Experimental Setup}

RQ2 analyzes the consistency component $C$, source prioritization $P$,
and confidence $c$ of the structured trace
$r_i^{v,m}=\langle A,C,P,H,\psi,c\rangle$ defined in
Section~\ref{sec:formula}. We evaluate four outcomes.

First, \emph{explicit conflict detection} measures whether a trace flags
a Javadoc--MUT inconsistency under \textsc{DocBug}, \textsc{MutBug}, and
\textsc{Conflict}, relative to the clean-sample false-positive rate. We
use IR-strict as the primary metric: it is the strict consolidated
inconsistency-report signal based on $C_{\mathrm{IR}}$, and counts a
case only when the trace explicitly reports a Javadoc--MUT conflict.
PCA, IC, Union, and Majority provide complementary views.

Second, \emph{source-prioritization discrimination} measures whether the
faulty artifact is ranked as less reliable. We compute Kendall's
$\tau_b$ between the provenance labels and the model's source ranking,
where positive values indicate correct deprioritization. Third,
\emph{description fidelity} is the cosine similarity between generated
conflict descriptions and the ground-truth fault annotation, computed
using \texttt{BAAI/bge-base-en-v1.5}. Fourth, \emph{confidence
calibration} measures the separation between confidence on detected and
missed conflicts.

This covers 456 samples per model per perturbed dataset
and is stratified by fault origin, contradiction strategy, and severity.\\

\begin{figure}[!t]
  \centering
  \includegraphics[width=\linewidth]{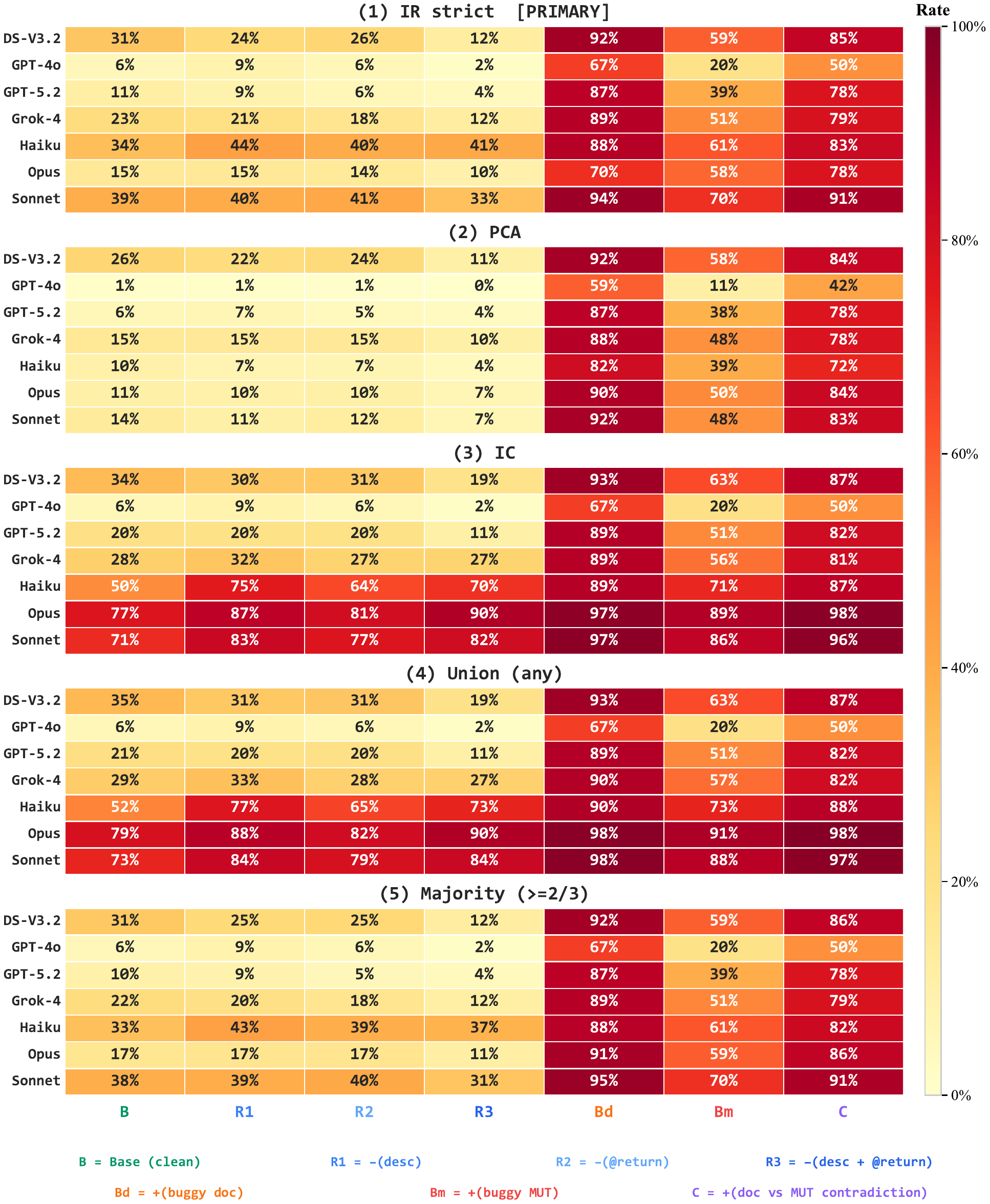}
  \caption{Javadoc--MUT inconsistency detection by perturbation type,
model, severity tier, and detection signal. Dashed lines show each
model's clean-sample false-positive rate; the main comparison is net
detection gain above this baseline.}
  \label{fig:rq2_detection_accuracy}
\end{figure}

\subsubsection{Results}

\emph{Conflict detection depends strongly on fault origin.}
Figure~\ref{fig:rq2_doc_vs_code_asymmetry} shows the central RQ2 result:
models detect Javadoc faults far more reliably than implementation
faults. After accounting for each model's clean false-positive rate,
Javadoc faults yield net detection gains of $+54.8$--$+76.5$~pp,
compared with only $+14.0$--$+42.5$~pp for MUT faults. Detection remains
close to \textsc{DocBug} performance when the Javadoc alone or both
artifacts are changed, with median differences of $+4.3$ and $+3.9$~pp
relative to \textsc{DocBug}, respectively. In contrast, when only the
MUT changes while Javadoc remains intact, detection falls by
$21$--$43$~pp relative to the both-changed condition. This \textit{asymmetry}
appears across \textit{all} seven models and \textit{all} five detection signals.

\emph{False-positive baselines are essential for interpreting detection.}
As Figure~\ref{fig:rq2_detection_accuracy} shows, clean-sample
false-positive rates range from $6\%$ for \texttt{GPT-4o} to $39\%$ for
\texttt{Sonnet}. Raw detection rates therefore reward models that flag
conflicts aggressively. For example, \texttt{GPT-5.2} detects $87.5\%$
of Javadoc faults, but achieves the largest net gain ($+76.5$~pp)
because its baseline is only $11\%$. On \textsc{Conflict}, IR-strict
detection ranges from $49.6\%$ for \texttt{GPT-4o} to $91.2\%$ for
\texttt{Sonnet}.

Union increases recall but sharply raises false positives. For
\texttt{Opus}, detection rises from $78.1\%$ under IR-strict to
$98.5\%$ under Union, while its clean false-positive rate rises from
$15.1\%$ to $78.9\%$. Majority closely tracks IR-strict. These results
support IR-strict as the primary metric because it requires explicit
conflict localization rather than broad suspicion.

\emph{Detecting a conflict does not ensure correct trust allocation.}
Only \texttt{Sonnet} ($\tau_b=0.33$) and \texttt{Opus}
($\tau_b=0.30$) meaningfully deprioritize faulty Javadocs. No model
reliably deprioritizes faulty implementations ($\tau_b\leq0.10$).
Discrimination improves for heavy Javadoc faults ($0.08$--$0.63$), but
approaches zero or becomes negative for subtle faults. \textit{Thus, even when
models report a mismatch, they often fail to lower trust in the
responsible implementation}.

\emph{Detected conflicts are described more faithfully, but confidence
rarely exposes failure.}
Detected traces align more closely with the ground-truth fault
descriptions than missed traces, with mean cosine similarity of $0.833$
versus $0.730$ and a gap of $0.104$
(Figure~\ref{fig:rq2_cosine_similarity}). This association holds across
all seven models, with \texttt{Opus}, \texttt{Sonnet}, and
\texttt{GPT-5.2} producing the most faithful descriptions. However,
Figure~\ref{fig:rq2_confidence_calibration} shows that only
\texttt{DeepSeek-V3.2-Speciale} meaningfully separates detected from
missed cases (gap $=0.73$); the remaining six models show little
separation.

\emph{Practical implication.}
Current LLMs can support first-pass Javadoc triage, subject to
model-specific false-positive rates. They should not be used as
stand-alone code-drift detectors: implementation-only conflicts remain
underdetected and require human review or complementary static, dynamic,
or symbolic analysis. Reported confidence is also unsuitable as an
automated acceptance gate without model-specific recalibration. RQ3 returns to this recognition-versus-localization distinction under
severity-controlled contradictions.

\begin{rqanswer}{RQ2 Findings}
LLMs are \textbf{asymmetric} consistency checkers. Javadoc faults produce net
detection gains of $+54.8$--$+76.5$~pp, compared with
$+14.0$--$+42.5$~pp for implementation faults; when only the MUT
changes, detection falls by $21$--$43$~pp. \textbf{No model reliably}
\textbf{deprioritizes} faulty implementations ($\tau_b\leq0.10$), and confidence
distinguishes detected from missed conflicts for only one of seven
models. \textbf{Current LLMs are therefore useful Javadoc auditors, but
unreliable stand-alone code-drift detectors}.
\end{rqanswer}

\subsection{RQ3: Semantic Code Understanding and Trace Fidelity}
\label{sec:rq3}

RQ1 and RQ2 reveal a common pattern: models respond more strongly to
documentation faults than to code faults, and detection drops when only
the MUT diverges. One possible explanation is reliance on explicit
lexical or structural cues rather than comparison of documented and
implemented behavior. RQ3 tests this explanation behaviorally by asking
whether detection and description fidelity remain stable as
Javadoc--MUT contradictions become less explicit. Smaller
heavy-to-subtle drops are consistent with stronger semantic comparison;
larger drops are consistent with greater reliance on surface evidence.\\

\subsubsection{Experimental Setup}

RQ3 analyzes the consistency component $C$ of the structured trace
$r_i^{v,m}=\langle A,C,P,H,\psi,c\rangle$. We use the full
\textsc{Conflict} dataset, stratified into the heavy, normal, and subtle
tiers defined in Section~\ref{subsec:perturbation_variants}. These tiers
progress from explicit behavioral conflicts to corner-case differences
requiring closer comparison of the Javadoc and implementation.

We measure two complementary outcomes. First, \emph{contradiction
detection} records whether IR-strict, the strict inconsistency-report
criterion based on $C_{\mathrm{IR}}$, identifies a Javadoc--MUT conflict.
Second, \emph{description fidelity} measures the cosine similarity
between the generated conflict description and the annotated fault
summary using \texttt{BAAI/bge-base-en-v1.5}. We report fidelity for
pairwise conflict analysis (PCA), explicit conflict lists (IC),
inconsistency reports (IR), and their combined representation. Because
missing IC or IR descriptions receive zero similarity, these measures
capture end-to-end trace fidelity: whether the model both surfaces and
accurately describes the conflict.

For each outcome, we use the heavy-to-subtle drop as the robustness
measure:
\[
\Delta_{\mathrm{det}}=\mathrm{Det}_{H}-\mathrm{Det}_{S},
\qquad
\Delta_{\mathrm{fid}}=\mathrm{Sim}_{H}-\mathrm{Sim}_{S}.
\]
Smaller drops indicate that performance remains stable as explicit cues
weaken; larger drops are consistent with greater reliance on surface
evidence.

\begin{figure}[!t]
  \centering
  \includegraphics[width=\linewidth]{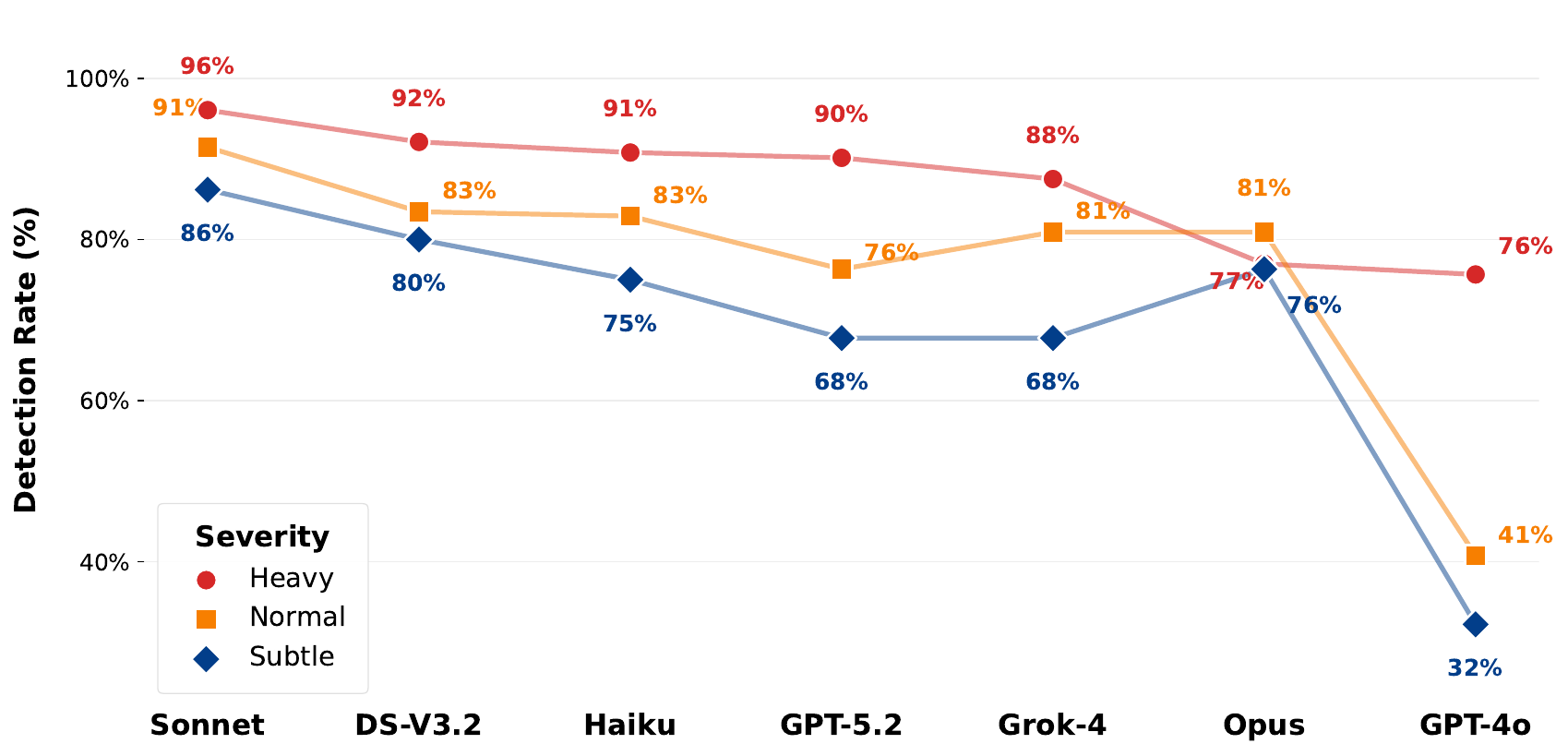}
  \caption{Severity-stratified Javadoc--MUT contradiction detection under
IR-strict evaluation ($n=152$ per severity tier). Sonnet is most robust
($96\%\rightarrow91\%\rightarrow86\%$), followed by DS-V3.2
($92\%\rightarrow83\%\rightarrow80\%$) and Haiku
($91\%\rightarrow83\%\rightarrow75\%$). GPT-4o shows the steepest drop
($76\%\rightarrow41\%\rightarrow32\%$). Opus is nonmonotonic
($77\%\rightarrow81\%\rightarrow76\%$) because many traces recognize the
mismatch but fail the artifact-localization requirement.}
  \label{fig:rq3_severity_detection_rate}
\end{figure}

\begin{figure*}[!t]
  \centering
  \includegraphics[width=0.6\linewidth]{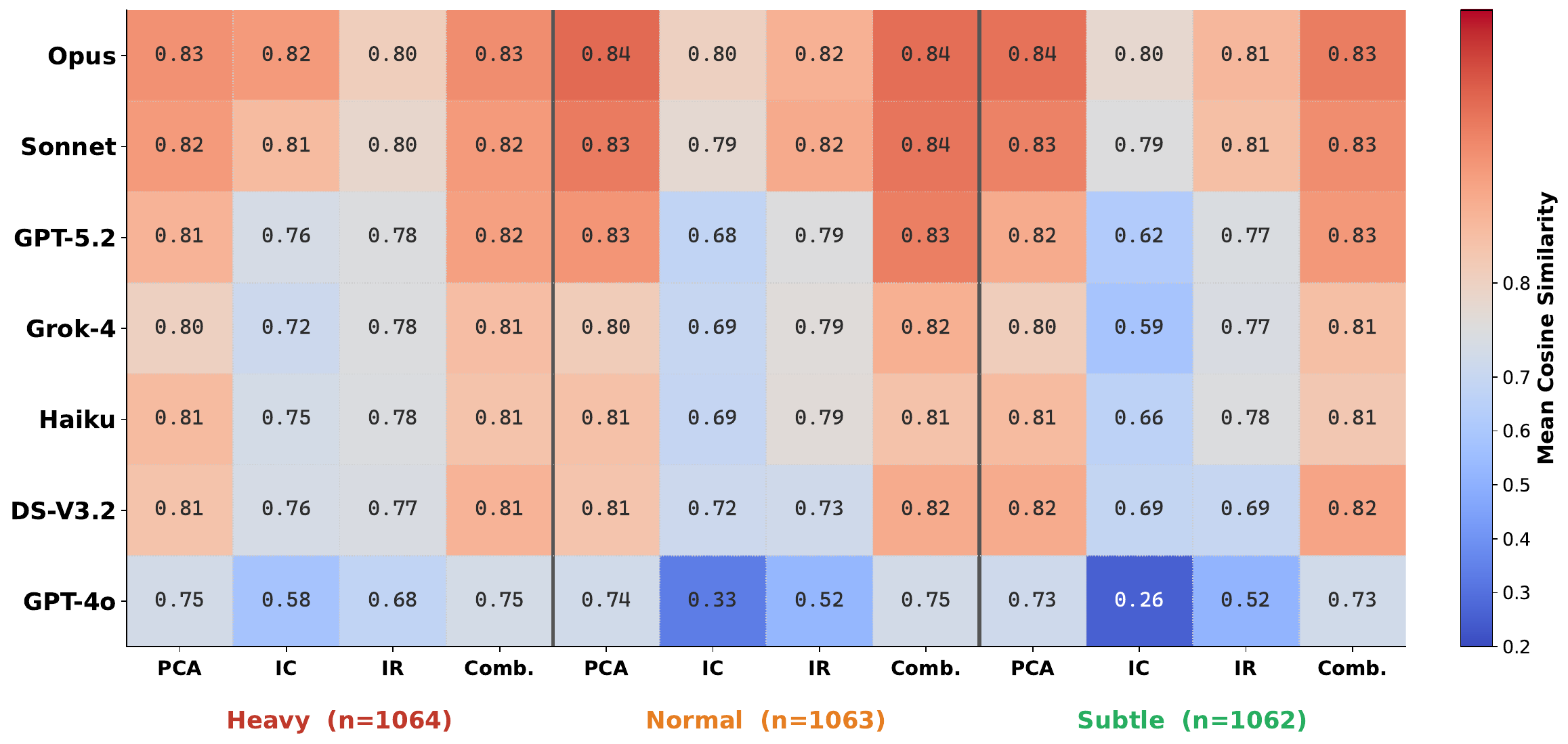}
  \caption{Description fidelity by severity for \textsc{Conflict},
measured as cosine similarity to ground-truth fault descriptions using
\texttt{BAAI/bge-base-en-v1.5}. Scores are shown for PCA, IC, IR, and
their combined representation. Missing IC or IR descriptions receive zero
similarity, so the metric reflects end-to-end trace fidelity.}
  \label{fig:rq3_cosine_by_severity}
\end{figure*}

\subsubsection{Results}

\emph{Subtle contradictions sharply separate the models.}
Figure~\ref{fig:rq3_severity_detection_rate} shows that Sonnet is the
most stable, declining from $96\%$ on heavy conflicts to $86\%$ on
subtle conflicts ($\Delta=10$~pp). DS-V3.2
($92\%\rightarrow83\%\rightarrow80\%$, $\Delta=12$~pp) and Haiku
($91\%\rightarrow83\%\rightarrow75\%$, $\Delta=16$~pp) are similarly
robust. GPT-5.2 and Grok-4 form an intermediate group, both reaching
$68\%$ on subtle conflicts, with drops of $22$ and $20$~pp,
respectively. GPT-4o shows the clearest collapse, falling from $76\%$
to $32\%$ ($\Delta=44$~pp). Thus, models that perform similarly on
explicit contradictions diverge substantially once the behavioral
difference becomes subtle.

\emph{Opus exposes the difference between recognition and localization.}
Opus is the exception to the severity trend
($77\%\rightarrow81\%\rightarrow76\%$), but the exception is
informative. Its traces often recognize the semantic mismatch while
identifying only the Javadoc as affected and omitting the MUT. Because
IR-strict requires the trace to report a Javadoc--MUT conflict, these
cases count as misses. Opus therefore shows that recognizing a
contradiction is not enough: the model must also localize the affected
artifacts correctly.

\emph{Description fidelity reinforces the same model separation.}
Figure~\ref{fig:rq3_cosine_by_severity} shows the largest differences in
the IC descriptions. GPT-4o declines from $0.58$ on heavy conflicts to
$0.33$ on normal and $0.26$ on subtle conflicts. In contrast, Opus
remains stable at $0.82\rightarrow0.80\rightarrow0.80$, and Sonnet at
$0.81\rightarrow0.79\rightarrow0.79$. Their combined similarities are
also stable: $0.83\rightarrow0.84\rightarrow0.83$ for Opus and
$0.82\rightarrow0.84\rightarrow0.83$ for Sonnet. PCA varies less across
severity, suggesting that pairwise conflict judgments are easier to
preserve than detailed fault characterization.

\emph{The surface-cue hypothesis receives a qualified answer.}
GPT-4o's steep decline in both detection and description fidelity as
contradictions become subtle is consistent with reliance on explicit
cues. Sonnet, Haiku, and DS-V3.2 remain much more stable as those cues
weaken, which is consistent with stronger semantic comparison of
Javadoc and implementation behavior. This explanation is incomplete,
however. As RQ2 showed, DS-V3.2 remains stable across severity tiers but
still exhibits a sizable implementation-only detection gap. Opus shows
the complementary failure mode: it often recognizes the mismatch but
mislocalizes the affected artifact. Thus, semantic robustness improves
trace fidelity, but source-origin bias and localization errors remain
separate failure modes.

\emph{Practical implication.}
Within this benchmark, Sonnet, DS-V3.2, and Haiku are the strongest
choices when workflows must detect subtle implementation-level
conflicts. GPT-4o is less reliable when code drift lacks explicit cues.
Opus produces consistently faithful descriptions, but may mislocalize
the responsible artifact; it is therefore better suited as a secondary
explainer than as a stand-alone detector.

\begin{rqanswer}{RQ3 Findings}
Models differ sharply as contradictions become subtle. Sonnet, DS-V3.2,
and Haiku retain comparatively stable detection, with heavy-to-subtle
drops of $10$--$16$~pp, whereas GPT-4o drops by $44$~pp. GPT-4o's
collapse is consistent with reliance on explicit cues, while Sonnet and
Opus preserve high description fidelity. Opus, however, frequently
mislocalizes the affected artifact. These results are consistent with
semantic comparison contributing to trace fidelity, but they do not
eliminate the implementation-only blind spot or the need to evaluate
localization separately.
\end{rqanswer}

\subsection{Threats to Validity}

\paragraph{Internal Validity.}
Some perturbations are LLM-generated, so samples labeled as
\emph{subtle} may still contain unintended cues. We mitigate this by
retaining only compilable, natural-looking variants and manually checking
generated samples for correct realization, contradiction validity, and
severity. Two authors manually checked the generated samples, and disagreements were
resolved through judgments.\\

\paragraph{External Validity.}
The benchmark contains 456 curated Java method bundles from 25 projects,
which may limit generalization to other languages, domains, or artifact
types. Controlled perturbations are also cleaner than real-world
inconsistencies, which are often partial, distributed, and semantically
entangled.\\

\paragraph{Construct Validity.}
\toolname{} quality scores measure perceived informativeness and internal
consistency, not ground-truth correctness. Kendall's $\tau_b$ orderings
derive from perturbation design rather than human rankings. Cosine
similarity with sentence-transformer embeddings is a proxy for
description fidelity, not a substitute for human evaluation.

\section{Related Work}

\paragraph{Robustness and prompt sensitivity.}
Empirical studies show that code LLMs are brittle to input variation. ReCode applies semantics-preserving perturbations to docstrings, identifiers, syntax, and formatting; NLPerturbator studies realistic natural-language variation; EMPICA contrasts semantics-preserving and semantics-changing code transformations; and \emph{When Prompts Go Wrong} evaluates ambiguous, incomplete, and contradictory task descriptions~\cite{Wang2023ReCode,Chen2026NLPerturbator,Nguyen2025EMPICA,Larbi2026WhenPromptsGoWrong}. These studies show that input form and quality affect downstream correctness. Their unit of analysis, however, is the stability or correctness of the final output. \toolname{} instead examines evidence allocation: when multiple artifacts are available, does the model recognize which artifact degraded and adjust its trust accordingly?

\paragraph{Hallucinations and output reliability.}
Prior work has cataloged bugs and hallucinations in LLM-generated code, including prompt misinterpretation, missing corner cases, invented program elements, and failures to respect repository context~\cite{Tambon2025Bugs,Zhang2025Hallucinations}. These studies characterize defects in generated artifacts and motivate post-generation validation. \toolname{} targets an earlier failure mode: a model may produce a plausible---or even correct---output while relying on the wrong evidence. It therefore evaluates per-artifact quality judgments, conflict attribution, source ranking, and confidence before downstream outputs hide that decision.

\paragraph{Noisy artifacts and cross-artifact consistency.}
In software testing, documentation quality affects oracle generation, and learned or LLM-based generators remain vulnerable to incomplete or noisy context~\cite{Doc2OracLL,HossainNeural2023,togll}. METAMON and DocPrism detect code--documentation inconsistencies, while Dietrich et al.\ verify consistency across programming-exercise artifacts~\cite{Lee2025METAMON,Xu2025DocPrism,dietrich2025llm}. These approaches either measure downstream performance or determine whether artifacts disagree. \toolname{} instead evaluates how a model responds to disagreement: paired, provenance-controlled perturbations jointly measure artifact-quality sensitivity, conflict localization, source prioritization, and confidence under a blind, artifact-symmetric protocol.

\paragraph{Positioning.}
Prior work asks whether outputs remain correct under perturbation, whether generated code hallucinates, or whether artifacts are inconsistent. \toolname{} asks a distinct reliability question: \emph{which artifact does an LLM trust when software evidence conflicts?} Its source-origin design further tests whether equivalent conflicts are treated consistently when introduced through documentation rather than implementation. This exposes artifact-specific over-trust and under-scrutiny that output-only and inconsistency-detection evaluations cannot reveal.

\section{Discussion}
\label{sec:discussion}

\textit{\toolname{} produces useful but uneven artifact-level judgments.}
The positive result is that structured traces are informative: quality
penalties localize to the perturbed artifact, scores preserve severity
ordering, and detected conflicts are described more faithfully than
missed ones. For example, removing both the Javadoc description and
\texttt{@return} lowers Javadoc scores by $0.300$--$0.463$, while MUT
scores change by less than $0.020$. Thus, the traces capture more than
generic suspicion.

\textit{The main limitation is artifact asymmetry.}
Across models, Javadoc faults are easier to detect, score, and
deprioritize than implementation faults. Severity separation is much
larger for Javadoc bugs than for MUT bugs ($0.152$--$0.253$ vs.\
$0.049$--$0.123$), and detection drops by $21$--$43$ percentage points
when only the MUT changes while plausible Javadoc remains intact. No
model reliably deprioritizes faulty implementations
($\tau_b \leq 0.10$). Current LLMs are therefore not symmetric
integrators of software evidence; they are stronger Javadoc auditors
than implementation-drift detectors.

\textit{Final-output accuracy is insufficient for evaluating trust.}
\toolname{} does not treat a model's explanation or confidence as ground
truth. Each trace is an observable judgment record, and correctness is
evaluated against known perturbation provenance. A confident but wrong
attribution remains wrong. A model may generate a plausible test, patch,
or review comment while relying on the wrong artifact; evaluations
should therefore measure not only task success, but also which artifacts
the model trusted to reach that result.

\textit{LLM traces should support triage, not replace review.}
LLMs can support first-pass Javadoc triage, especially for stale,
incomplete, or contradictory documentation. They should not be used alone
to detect subtle implementation drift. When code behavior is the risk,
LLM traces should be paired with human review or programmatic checks
such as static analysis, dynamic testing, or symbolic reasoning.
Reported confidence should also not be used as an automated acceptance
gate: six of seven models show little separation between correct and
incorrect judgments.

\textit{\toolname{} extends beyond this Java testing benchmark.}
The same design can be instantiated over requirements, patches, tests,
documentation, review context, or agent memory, with perturbations chosen
for workflow-specific risks. Practitioners can use it as a predeployment
audit to identify over-trusted artifacts, missed conflicts, and review
needs. Researchers can use it to compare prompts, context-selection
strategies, fine-tuning methods, and agent designs on artifact-aware
behavior rather than downstream accuracy alone.

\textit{\toolname{} is a regression framework, not a leaderboard.}
Its purpose is to measure whether new models, prompts, or agents preserve
desired trust behavior as they evolve. Re-running the benchmark can
reveal behavioral drift, such as improved code scrutiny, increased
Javadoc anchoring, or degraded confidence reliability. The main takeaway
is that current LLMs can assist artifact triage, but correctness-critical
SE workflows still need explicit mechanisms for deciding what evidence
deserves trust.

\section{Conclusion}

\toolname{} asks a simple but often hidden question in LLM-based software
engineering: before a model generates a test, patch, review, or summary,
does it know which artifact to trust? We introduced \toolname{} as a
model-neutral controlled-perturbation framework for eliciting and
evaluating structured artifact-level trust traces. Each trace records
per-artifact quality judgments, cross-artifact inconsistency analysis,
affected-source attribution, source prioritization, and confidence; these
judgments are evaluated against known perturbation provenance rather than
accepted as self-reported truth.

We instantiated \toolname{} on Java unit-testing bundles by constructing
aligned base and perturbed versions of 456 real-world methods and
collecting 22,339 valid traces from seven LLMs. The results show useful
but uneven trust behavior. Models often localize degradation and track
severity, but they are far more sensitive to Javadoc faults than to
implementation faults. They reliably surface documentation-side errors,
yet miss many implementation-only drifts when plausible Javadoc remains
intact. They also fail to consistently deprioritize faulty
implementations, and confidence separates correct from incorrect
judgments for only one model. Robustness under subtle contradictions is
consistent with semantic code understanding contributing to trace
fidelity, but it does not remove the implementation-drift blind spot.

The broader lesson is that final-output correctness is not enough. LLM
assistants must also be evaluated on how they allocate trust across the
artifacts used to produce that output. \toolname{} provides a reusable
way to expose artifact-specific blind spots before deployment, compare
models and prompts beyond task accuracy, and decide where human review
or static, dynamic, or symbolic checks remain necessary. Current LLMs can
help triage Javadoc quality, but correctness-critical SE workflows still
need explicit safeguards for deciding what evidence deserves trust.

\section{Data Availability}

To support transparency and reproducibility, we will make the \toolname{} artifact, including the benchmark, prompts, and evaluation scripts, publicly available upon publication of this work.

\balance
\bibliographystyle{IEEEtran}
\bibliography{main}
\end{document}